\documentclass[aps,twocolumn,showpacs]{revtex4}
\usepackage{graphicx}
\usepackage{slashbox}

\begin{document}

\title{Spectral Line Width Broadening from Pair Fluctuation in a Frozen Rydberg Gas}

\author{B. Sun and F. Robicheaux}

\affiliation{Department of Physics, Auburn University, Auburn, AL
36849, USA}

\begin{abstract}
Spectral line width broadening in Rydberg gases, a phenomenon
previously attributed to the many-body effect, was observed
experimentally almost a decade ago. The observed line width was
typically 80-100 times larger than the average interaction
strength predicted from a binary interaction. The interpretation
of such a phenomenon is usually based on the so-called diffusion
model, where the line width broadening mostly originates from the
diffusion of excitations. In this paper, we present a model
calculation to show that diffusion is not the main mechanism to
the line width broadening. We find that the rare pair fluctuation
at small separation is the dominant factor contributing to this
broadening. Our results give a width of about 20-30 times larger
than the average interaction strength. More importantly, by
turning off the diffusion process, we do not observe order of
magnitude change in the spectral line width.
\end{abstract}


\maketitle

Rydberg gases have attracted renewed interest in recent years due
to the unprecedented advancement in laser cooling and trapping
\cite{gallagher,pillet,martin,raithel,gould,singer,pillet2,pfau,Heuvell}.
Rydberg atoms, possessing a large dipole moment and long lifetime,
can interact with each other coherently for relatively long times,
which make it a potential candidate for quantum information
processing \cite{lukin,zoller}. There have been many experiments
exploring the quantum many-body effects in Rydberg gases, e.g.
spectral line broadening \cite{gallagher,pillet,martin}, number
correlation \cite{raithel,rost,francis} and collective excitation
\cite{pfau}. Many experimental results can be understood from the
well-known dipole blockade effect: when two Rydberg atoms are
close enough, the dipolar interaction will shift them out of
resonance with the external driving laser, thus double excitation
is greatly suppressed.

In this paper, we are interested in the unusual line width broadening which was observed in experiments
\cite{gallagher,pillet}. To be specific, we will consider the following two cases. (I), the main process
is $|np\rangle+|np\rangle \leftrightarrow |ns\rangle+|(n+1)s\rangle$ for experiment \cite{pillet}, where
the principal quantum number $n$ is 23 and the maximal gas density is around $10^{10}{\rm cm}^{-3}$.
$np$, $ns$, and $(n+1)s$ are abbreviated as $p$, $s$, and $s'$, respectively. (II), the main process is
$|(n+1)s\rangle+|n's\rangle \leftrightarrow |np\rangle+|(n'+1)p\rangle$ for experiment \cite{gallagher},
where the principal quantum numbers $n$ and $n'$ are 24 and 33, respectively, and the maximal gas
density is around $10^9 {\rm cm}^{-3}$ for each of the $s$ state. $(n+1)s$, $n's$, $np$, and $(n'+1)p$
are abbreviated as $s$, $s'$, $p$, and $p'$, respectively. For both cases, they express the creation
process, e.g., in case (I), one atom makes a downward transition from the Rydberg state $|p\rangle$ to
$|s\rangle$, and the other atom makes an upward transition from $|s'\rangle$ to $|p\rangle$, that is to
say, creating $ss'$ from a $pp$ pair. The detuning between $|pp\rangle$ and $|ss'\rangle$ is controlled
by a static electric field and the transition is allowed with dominant dipole moments $\mu_{ps}$ and
$\mu_{ps'}$. Here $\mu_{\alpha\beta}$ denotes the transition dipole moment between states
$|\alpha\rangle$ and $|\beta\rangle$. Similar notations will apply to case (II). In addition to the
above creation processes, there also exist the exchange process, e.g.
$|p\rangle+|s\rangle\leftrightarrow |s\rangle+|p\rangle$. Different from creation process, the exchange
process is always resonant and it describes the hopping of excitation in the whole gas. For this reason,
we will also call it a diffusion process.

A rough estimate from binary interaction will give a line width of
the order of the average interaction strength
$\bar{V}_1=\mu_{ps}\mu_{ps'}n$ or $\bar{V}_2=\mu_{ps}\mu_{p's'}n$
for the two cases respectively, with $n$ the average density of
the gas. However, the experimentally observed line widths are
typically $\sim 100\bar{V}_1$ for the first case, and are $\sim 80
\bar{V}_2$ for the second case. Previous explanations are based on
a diffusion model, where resonant processes like
$|p\rangle+|s\rangle \leftrightarrow |s\rangle+|p\rangle$ form a
diffusion band and play a dominant role in the broadening. In this
model, even at large detuning, there are still some pairs of atoms
close enough to perform the creation process. Because the hopping
of excitation can happen in the whole gas, the diffusion evacuates
the excitations so that each pair of close atoms can react several
times, analogous to autocatalytic processes in chemistry. As
pointed out in Ref. \cite{pillet}, the band formed by the
diffusion of $|ss'\rangle$ is coupled to the state $|ss'\rangle$
and the existence of this band broadens this population transfer
\cite{pillet,akulin}, which shows that the broadening should be a
result of the many-body effect. However, we believe that it is
mostly a two-body effect arising from the density fluctuation, as
we will show below by simple theoretical reasoning and numerical
simulations.

\begin{figure}[htb]
\includegraphics[width=3.25in]{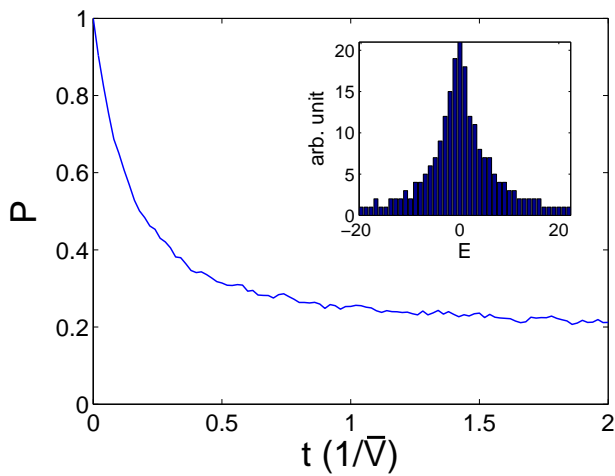}
\caption{The probability of finding $s$ state on the initial $s$ atom as function of time. The inset
shows the histogram of the eigen-energy. The simulation is carried out with 256 atoms and over 1000
spatial configurations. Time is in unit of $1/\bar{V}$ (see text).} \label{decayE}
\end{figure}

To better understand the diffusion process, we consider a toy
model for the process $|s\rangle+|p\rangle\leftrightarrow
|p\rangle+|s\rangle$. The Hamiltonian is
\begin{equation}
H=\sum_{j,k}V_{\rm dip}(\vec r_j-\vec r_k), \label{dh}
\end{equation}
which is purely a diffusion process under the dipolar interaction
\begin{equation}
V_{\rm dip}(\vec r)=c_d{1-3\cos^2\theta \over r^3}.
\end{equation}
$\vec r=\vec r_1-\vec r_2$, $\cos\theta=\hat{z}\cdot \hat{r}$, and $c_d=\mu_{sp}^2$. The energy scale is
chosen to be $\bar{V}=\mu_{sp}^2 n$, the inverse of which sets up the time scale. To ease our
discussion, we consider only one $s$ atom and a bunch of $p$ atoms with zero magnetic moment, i.e.,
magnetic quantum number $m=0$. In Fig. \ref{decayE}, we show the probability of finding an $s$ state on
the initial $s$ atom as function of time, i.e., $P=|\langle \psi(t)|\psi(t=0)\rangle|^2$ where
$|\psi(t)\rangle$ is the many-body wave function at time $t$. We can see that the probability decays
smoothly and saturates to a finite value. At $t\sim 0.2$, there is about 50\% chance that the $s$ state
has drifted away. So the characteristic time of diffusion is on the order of 1/10. In addition, we show
the histogram of eigen-energy in the inset of Fig. \ref{decayE} by directly diagonalizing the
Hamiltonian (\ref{dh}). We find that the width of this diffusion band is roughly 5, corresponding to
$1/t$. This poses questions on the original explanation of the band diffusion model. Since for large
detuning, e.g. $\Delta=40 \bar{V}_1$, in order to make a non-negligible transfer from $pp$ to $ss'$, the
interaction strength between them should be of the same order as $\Delta$. In this case, $pp$ and $ss'$
should be split by an amount of $\sim \Delta$. That is to say, the manifold of $pp$ and $ss'$ will be in
a large detuning to the band. Thus, the state $ss'$ with a large detuning from the band will not decay
into it. Therefore, the explanation of broadening from this diffusion band model is questionable.

We investigate the line width problem of case (I) and (II) by
direct numerical simulations. To do this, we randomly put several
atoms in a cubic box and assume each atomic state $|p\rangle$ or
$|p'\rangle$ has no magnetic moment. The numerical schemes are as
follows. We first calculate the full dynamics with a given
detuning for each spatial configuration of atoms with a fixed
evolution time. We then average over spatial configurations to
obtain the excitation probability as function of detuning from
which we can extract the line width. In order to minimize the
finite size effect, we adopt wrap boundary condition.

We first focus on case (I) and discuss case (II) subsequently. The
energy and time scale are the same as those in the toy model. In
case (I), each atom can be in the states $p$, $s$, or $s'$. The
Hamiltonian is found to be
\begin{eqnarray}
H &=& \sum_{jk} [ V_{jk} e^{-i\Delta t}|p_jp_k\rangle\langle
s_js'_k|+
V'_{jk} |p_js_k\rangle\langle s_jp_k|  \nonumber\\
&+& V''_{jk} |p_js'_k\rangle\langle s'_jp_k| ]+h.c. \nonumber
\end{eqnarray}
describing the following processes
\begin{eqnarray}
p+p &\to& s+s' ,    \label{p1}\\
p+s &\to& s+p ,\label{p2}\\
p+s' &\to& s'+p \label{p3},
\end{eqnarray}
where process (\ref{p1}) is not always resonant and its detuning
$\Delta$ is controlled by an electric field, while processes
(\ref{p2}) and (\ref{p3}) are always resonant. $V_{jk}$, $
V'_{jk}$, and $V''_{jk}$ all take the form of $V_{\rm dip}(\vec
r_j-\vec r_k)$ with corresponding
$c_d=\mu_{sp}\mu_{sp'},\mu_{sp}^2,$ and $\mu_{sp'}^2$ for
processes (\ref{p1}), (\ref{p2}) and (\ref{p3}), respectively.
Initially all of the atoms are in the state $p$, and they evolve
under the dipolar interaction for a fixed time $T$. We are
interested in the yield of $s$ atom ($f_s$) as function of
detuning, from which we can extract the line width.

\begin{figure}[htb]
\includegraphics[width=3.25in]{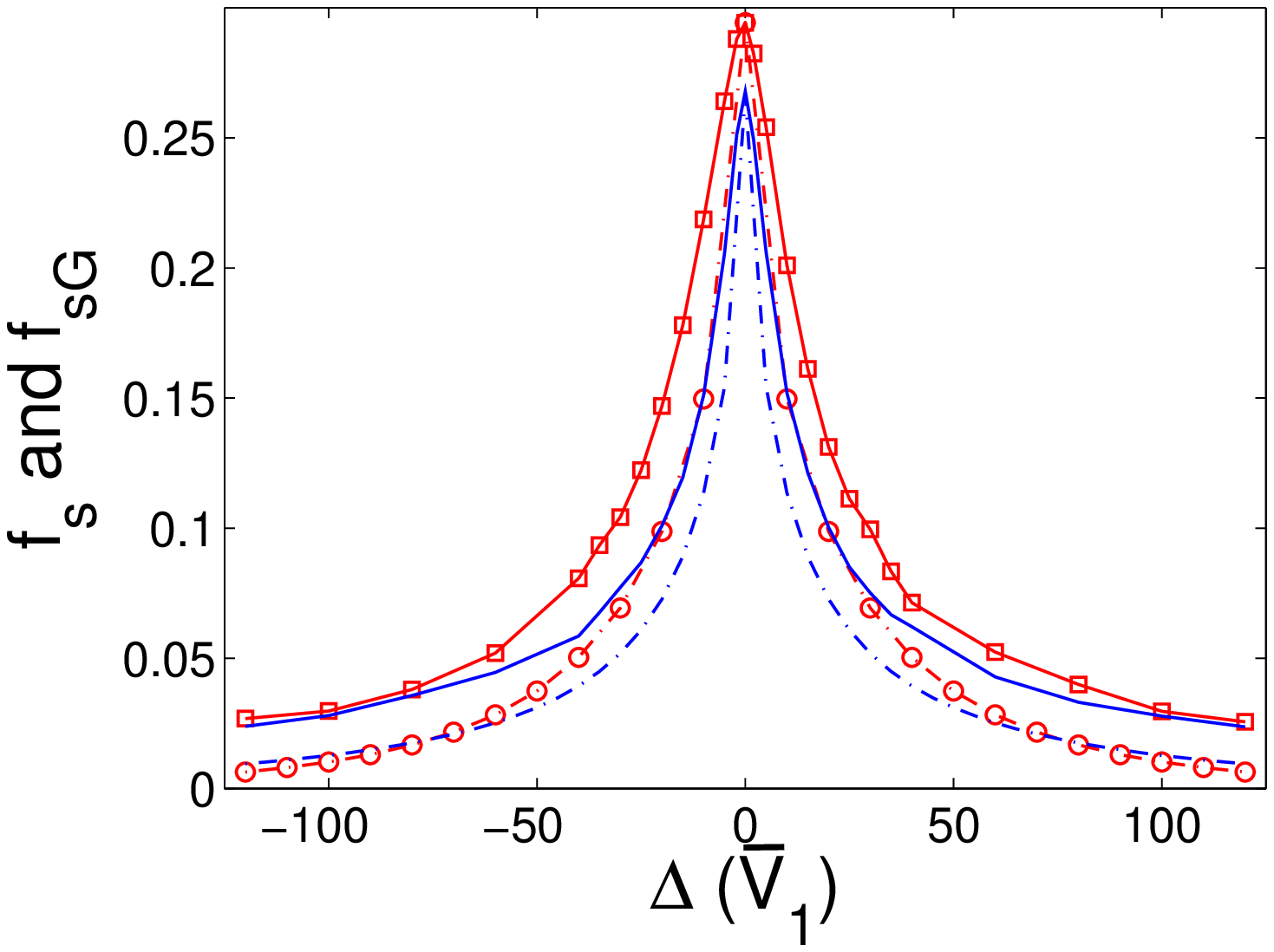}
\caption{(Color online)$f_s$ and $f_{s,G}$ as function of $\Delta$ for $N=10$ atoms averaged over 1000
spatial configurations. Results with marked (unmarked) lines include (exclude) diffusion. Solid and dash
dotted lines are for the case of homogeneous and Gaussian convolution, respectively. Parameters used:
$T=3.4$, $\sigma=500$, $\mu_{sp}=1.02,\mu_{sp'}=0.98$. The results are insensitive to $\sigma$. $\Delta$
is in unit of $\bar{V}_1$ (see text).} \label{10}
\end{figure}

We perform calculation with up to $N=10$ atoms and average over
the initially random atom positions many times. Under wrap
boundary condition, our results already show convergent behavior
for $N=8$ atoms. Extrapolation to $N=\infty$ will give about
$15\%$ difference to that of $N=10$. As we will see later, this
difference is not crucial to our conclusion as we are interested
in the order of magnitude difference. Our results are shown as
solid lines in Fig. \ref{10}. For the marked solid line, we
include all three processes, while for the unmarked solid line, we
only consider the process (\ref{p1}), i.e., we effectively turn
off the diffusion process. The extracted line widths are about 35
and 25 for the two cases, respectively. This shows that the
process (\ref{p1}) already gives a width of order of magnitude
larger than the average interaction strength and the diffusion
process further broadens the line width by roughly 50\%. So the
unusual line width broadening mostly comes from the process
(\ref{p1}).

In real experiments, the density is not uniform. Therefore, we
need to take into account the density profile of the atomic cloud.
We assume it as a Gaussian form with a width $\sigma$, so the
density is written as $n(r) = 2\sqrt{2}\bar{n}e^{-r^2/\sigma^2}$
where $\bar{n}$ is the average density. Accordingly, the
excitation fraction with a Gaussian convolution ($f_{s,G}$) is
given by
\begin{eqnarray}
f_{s,G} &=& {\int f_s(\Delta/\bar{V}\times \bar{V}/V(r)) n(r) r^2dr \over
                \int n(r) r^2 dr}   \nonumber\\
         &=& {\int f_s(\Delta/\bar{V}\times e^{r^2/\sigma^2}/2\sqrt{2})
         e^{-r^2/\sigma^2}  r^2dr
                \over \int e^{-r^2/\sigma^2}  r^2 dr}   \nonumber\\
\end{eqnarray}
Here we assume the density varies slowly on the length scale we
considered. The results for $N=10$ atoms are shown as the
dash-dotted lines in Fig \ref{10}. The marked dash-dotted line is
the result including all three processes. While the unmarked
dash-dotted line is the result only including process (\ref{p1}).
The extracted line width with diffusion is about 30, which is a
few times smaller than the results in Ref. \cite{pillet}. The one
extracted from the calculation without diffusion is about 20,
again demonstrating that diffusion is not the main mechanism in
the broadening. We also note that the curves are not perfectly
symmetric around $\Delta=0$ due to anisotropy of the dipolar
interaction. However, this difference is too small to be detected
under current experimental conditions.

We further consider the motional effect on line width broadening.
To do this, we give a constant speed $v_s$ for each atom but with
random direction. For $v_s=0.05$, the atom moves $0.1$ (average
distance) at the end of simulation,which is still in the so-called
``frozen gas" regime. Our results give an additional broadening of
about $20\%$, so the motional effect is not important as expected.

\begin{figure}[htb]
\includegraphics[width=3.25in]{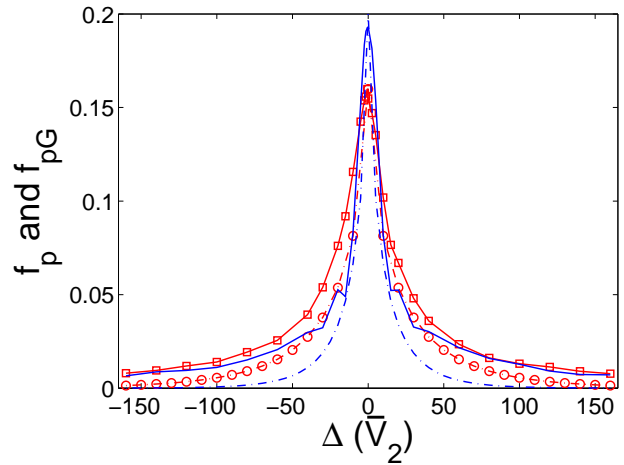}
\caption{(Color online)$f_p$ and $f_{p,G}$ as function of $\Delta$ for a total $N=20$ atoms averaged
over 1000 spatial configurations. Results with marked (unmarked) lines include (exclude) diffusion.
Solid and dash dotted lines are for the case of homogeneous and Gaussian convolution, respectively.
Parameters used: $T=0.36$, $\sigma=500$, $\mu_{sp}=2,\mu_{s'p'}=0.5$. The results are insensitive to
$\sigma$. The initial population of $s$ and $s'$ are the same. $\Delta$ is in unit of $\bar{V}_2$ (see
text).} \label{10g}
\end{figure}

Following similar procedures, we investigate case (II). Now each
atom can be in the states $s$, $s'$, $p$, or $p'$. The Hamiltonian
is found to be
\begin{eqnarray}
H &=& \sum_{jk} [ V_{jk} e^{-i\Delta t}|p_jp'_k\rangle\langle
s_js'_k|+
V'_{jk} |p_js_k\rangle\langle s_jp_k|  \nonumber\\
&+& V''_{jk} |p'_js'_k\rangle\langle s'_jp'_k| ]+h.c. \nonumber
\end{eqnarray}
describing the following processes
\begin{eqnarray}
s+s' &\to& p+p' ,    \label{p1a}\\
p+s &\to& s+p ,\label{p2a}\\
p'+s' &\to& s'+p' \label{p3a},
\end{eqnarray}
with dipolar interaction. Our numerical results are shown in Fig.
\ref{10g}. In this case, the width with (without) diffusion is
found to be about 20 (10). Therefore, the diffusion does not cause
order of magnitude change in the line width broadening.

So what causes the broadening? It is nothing unusual but the rare
pair fluctuation at small distances. To see this point, we
calculate $P(|\Delta|)$, which is the probability distribution of
nearest neighboring atoms with the absolute interaction strength
not larger than the absolute detuning. $P(|\Delta|)$ can be found
with the help of
\begin{equation}
P(|\Delta|)=\int_0^{|\Delta|} {dP(|V|\le|\Delta'|) \over
d|\Delta'|} d|\Delta'|
\end{equation}

In a homogeneous system, according to the Erlang distribution \cite{erlang}, the nearest pair
distribution is $\propto e^{-4\pi r^3/3}$ with unit density. Therefore, for an isotropic interaction
$V_{\rm iso}=1/r^3$ with $c_d=1$ (e.g. $\bar{V}=1$), we have
\begin{equation}
{dP(|V_{\rm iso}|\le|\Delta|) \over d|\Delta|}={4\pi\over 3
|\Delta|^2 }e^{-{4\pi \over 3|\Delta|}}.
\end{equation}
While for dipolar interaction $V_{\rm dip}$ with $c_d=1$ (e.g. $\bar{V}=1$), we have
\begin{equation}
{dP(|V_{\rm dip}|\le|\Delta|) \over d|\Delta|}={4\pi\over 3
|\Delta|^2 }\int_0^1 dx |1-3x^2| e^{-{4\pi \over
3|\Delta|}|1-3x^2|} .
\end{equation}

As $|\Delta|\rightarrow +\infty$, the asymptotic behaviors are
$(4\pi/3)|\Delta|^{-2}$ and $(16\pi/9 \sqrt{3})|\Delta|^{-2}$ for
isotropic and dipolar interaction, respectively. As
$|\Delta|\rightarrow 0^+$, they approach $0$ and $\sqrt{3}/(4\pi)$
for isotropic and dipolar interaction, respectively. The
remarkable difference at $|\Delta|\rightarrow 0^+$ is a signature
of the dipolar interaction.

\begin{figure}[htb]
\includegraphics[width=3.25in]{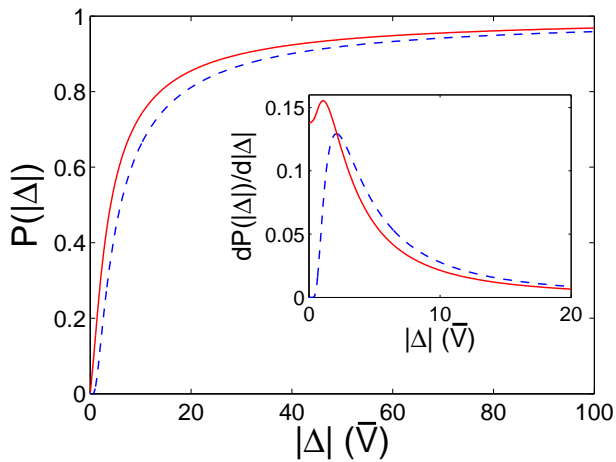}
\caption{(Color online)$P(|\Delta|)$ as function of $|\Delta|$. Blue dashed lines are for isotropic
interaction and red solid lines are for anisotropic interaction. $\Delta$ is in unit of $\bar{V}$ (see
text).} \label{prob}
\end{figure}

For $|\Delta|=40$, the probability of nearest atom pairs that has
an interaction energy larger than $|\Delta|$ is about 10\%, i.e.
those atom pairs will have non-negligible contribution to the
dynamics. The calculated $f_s$ of case (I) at this detuning is
about 8\%, close to the estimated value. Therefore, the rare pair
fluctuation is the main cause of the line width broadening and the
diffusion of excitation further increases this broadening by
roughly $50\%$.

\begin{figure}[!htb]
\includegraphics[width=3.25in]{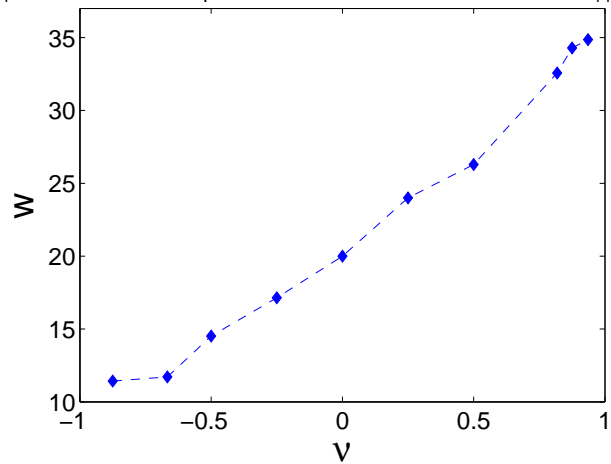}
\caption{Line width $w$ as function of population ratio $\nu$ of s and s' atoms. The same parameters are
used as in Fig. \ref{10g} except for different ratio of $s$ and $s'$ atoms.} \label{width}
\end{figure}

For case (II), we can also compute the line width ($w$) for different ratio of $s$ and $s'$, i.e. $w$ as
function of $\nu\equiv (n_1-n_2)/(n_1+n_2)$ where $n_1 (n_2)$ is the density of $s (s')$ atoms. Our
numerical results are shown in Fig. \ref{width}. The largest error in our simulation still comes from
the finite atom effect which has been discussed for Figs. 2 and 3. This error does not change much as we
vary the ratio of $s$ and $s'$ atoms. Other errors are negligible. We find that $w$ increases as $\nu$
increases and saturates at $\nu=\pm 1$. The increasing behavior of $w$ with $\nu$ is due to the
imbalanced hopping of s and s' atoms ($\mu_{sp}=4\mu_{s'p'}$). The increased ratio in s atoms will thus
show a stronger diffusion effect. However, this increase in the line width is again not an order of
magnitude change.

To conclude, we have reexamined the important role that pair
fluctuations play in the spectral line width broadening of a
frozen Rydberg gas. From direct numerical simulations, we find
that density fluctuations contribute to a width of roughly $20-30$
times of average interaction strength. In addition, by turning off
the diffusion process, we did not find order of magnitude change
in the line width. Therefore, the large line width is primarily
due to density fluctuations, and the diffusion process is not
overwhelmingly dominant as previously suggested. However, our
numerical results are only in qualitatively agreement with
experimental results. The even larger line width observed in both
cases and the double peak structure as observed in (I) cannot be
explained from current calculations which encourages further
investigations.

This work is supported by the NSF under grant no. 0653301.

\end{document}